\pdfoutput=1
\RequirePackage{snapshot}
\documentclass[sigconf,screen]{acmart}
\usepackage{etex} % MUST come right after class - fixes 'out of \dimen' error
\PassOptionsToPackage{colorlinks=true,citecolor=violet,urlcolor=blue}{hyperref}
\PassOptionsToPackage{protrusion=true,expansion=true,kerning,spacing}{microtype}
\PassOptionsToPackage{sort}{natbib}
\PassOptionsToPackage{usenames,dvipsnames,svgnames}{xcolor}

\usepackage{adjustbox}    % Auto-resize table content (eg Opo SenSys'14 rel)
\usepackage{amsfonts}     % Adds math fonts, commands such as \begin{align}
\usepackage{amsmath}      % Provides align* environment
\usepackage{array}        % Tables for use in math mode
\usepackage{balance}      % For balanced columns on the last page
\usepackage{booktabs}     % Elegant table-formatting library
\usepackage{bold-extra}   % Provides bf+sc (only in textbf+textsc env.)
\usepackage{bytefield}    % Formatting and layout of packets / bytefields
\usepackage[skip=5pt]{caption}
\usepackage{colortbl}     % Color table cells
\usepackage{comment}      % Provides \begin,\end{comment} for large blocks
\usepackage{cprotect}     % Allows verbatim, other formatting in macro args
\usepackage[ampersand]{easylist} % Simpler list syntax
\usepackage{endnotes}     % Footnotes pushed to the end of a document
\usepackage{enumitem}     % Allow custimizations of itemize/enumerate env's
\usepackage{float}        % Allow use of [H] to force figure placement
\usepackage{framed}       % codefigures
\usepackage{gensymb}      % Adds useful symbols w/out math mode, e.g. \degree
\usepackage{graphicx}     % For importing graphics
\usepackage{lipsum}       % For generating temporary filler text
\usepackage{listings}     % in-line source code (poorly, consider minted)
\usepackage{marginnote}   % ... for margin notes
\usepackage{mathtools}    % amsmath extension, adds more math formatting
\usepackage[protrusion=true,expansion=true,kerning,spacing]{microtype} % better type, spacing
\usepackage{multirow}     % Multiple row spacing in tables
\usepackage{nth}          % Typeset 33rd correctly as \nth{33}
\usepackage{rotating}     % Rotates any object, note sideways != sidewaysfigure
\usepackage{soul}         % Provides \hl{} for highlighting
\usepackage[subrefformat=parens]{subcaption}   % Replaces both subfig and subfigure
\usepackage{tabularx}     % Complicated table creation
\usepackage{tcolorbox}    % Colored boxes
\usepackage{textcomp}     % Provides \textmu for upright mu's
\usepackage{threeparttable} % Add footnotes to a table
\usepackage{tikz}         % okay, this is just for a pretty checkmark
\usepackage{units}        % For nice fractions, \nicefrac{1}{2} --> 1/2
\usepackage{url}          % Pretty printing of hyperlinks
\usepackage[usenames,dvipsnames,svgnames]{xcolor} % Allow the use and definition of colors
\usepackage{xspace}       % Intelligently add spaces after commands
\usepackage{xfrac}

\usepackage[colorlinks=true,citecolor=violet,urlcolor=blue]{hyperref}     % Creates hyperlinks from ref/cite
\hypersetup{pdfstartview=FitH} % Sets default zoom to 100% width

\usepackage[capitalise,nameinlink,noabbrev]{cleveref}     % Do the right thing with fig/table references

\setcopyright{none}
\acmDOI{PleaseReplaceWithProperDOI}
\acmISBN{Please replace with IEEE ISBN}
\acmYear{2018}
\copyrightyear{2018}
\acmPrice{}
\acmConference[IPSN'18]{The 17th ACM/IEEE Conference on Information Processing in Sensor Networks}{April 11-13, 2018}{Porto, Portugal}

\newlength\SUBSIZE

\makeatletter
\newwrite\remember@figures
\AtBeginDocument{%
  \InputIfFileExists{\jobname.dft}{}{}%
  \immediate\openout\remember@figures=\jobname.dft
}
\AtEndDocument{\immediate\closeout\remember@figures}

\newcommand{\placefigure}[2][tp]{%
    \csname remembered@figure@#2\endcsname{#1}
}

\NewEnviron{dfigure}[1]{%
  \immediate\write\remember@figures{%
    \noexpand\rememberfigure{#1}{\unexpanded\expandafter{\BODY}}%
  }%
}
\NewEnviron{dfigure*}[1]{%
  \immediate\write\remember@figures{%
    \noexpand\rememberfigurestar{#1}{\unexpanded\expandafter{\BODY}}%
  }%
}
\NewEnviron{dtable}[1]{%
  \immediate\write\remember@figures{%
    \noexpand\remembertable{#1}{\unexpanded\expandafter{\BODY}}%
  }%
}
\NewEnviron{dtable*}[1]{%
  \immediate\write\remember@figures{%
    \noexpand\remembertablestar{#1}{\unexpanded\expandafter{\BODY}}%
  }%
}

\newcommand{\rememberfigure}[2]{%
  \global\@namedef{remembered@figure@#1}##1{%
    \begin{figure}[##1]#2\label{#1}\end{figure}%
  }%
}
\newcommand{\rememberfigurestar}[2]{%
  \global\@namedef{remembered@figure@#1}##1{%
    \begin{figure*}[##1]#2\label{#1}\end{figure*}%
  }%
}
\newcommand{\remembertable}[2]{%
  \global\@namedef{remembered@figure@#1}##1{%
    \begin{table}[##1]#2\label{#1}\end{table}%
  }%
}
\newcommand{\remembertablestar}[2]{%
  \global\@namedef{remembered@figure@#1}##1{%
    \begin{table*}[##1]#2\label{#1}\end{table*}%
  }%
}
\makeatother

\definecolor{applegreen}{rgb}{0.55, 0.71, 0.0}
\definecolor{amethyst}{rgb}{0.6, 0.4, 0.8}
\definecolor{asparagus}{rgb}{0.53, 0.66, 0.42}
\definecolor{offwhite}{rgb}{1.0, 0.99, 0.97}

\graphicspath{{../figs/}{../images/}}

  \renewenvironment{thebibliography}[1]{%
    \begin{oldthebibliography}{#1}%
      \setlength{\parskip}{0ex}%
      \setlength{\itemsep}{0ex}%
  }%
  {%
    \end{oldthebibliography}%
  }

\newcommand{\uA}{{\textmu}A\xspace}

\newcommand{\iic}{I$^2$C\xspace}

\newcommand{\mm}{m\textsuperscript{2}\xspace}

\newcommand{\name}{Signpost\xspace}
\newcommand{\names}{Signposts\xspace}

\definecolor{light-gray}{gray}{0.75}

\def\checkmark{\tikz\fill[scale=0.4](0,.35) -- (.25,0) -- (1,.7) -- (.25,.15) -- cycle;} 

\makeatletter
\def\blfootnote{\xdef\@thefnmark{}\@footnotetext}
\makeatother

\newcommand\cczerotext{%
  {\footnotesize
    {\raisebox{-.3\height}{\includegraphics[width=.07\linewidth]{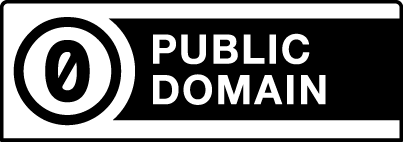}}}
    {
      To the extent possible under law, the authors have waived all copyright and
      related or neighboring rights to this work.
      This work is published from the United States.
    }
   }}
\newcommand\cczeronotice{%
\begin{tikzpicture}[remember picture,overlay]
  \node[anchor=south,yshift=30pt] at (current page.south) {\parbox{\dimexpr\textwidth-\fboxsep-\fboxrule\relax}{\cczerotext}};
\end{tikzpicture}%
}

\begin{document}

\title{The Signpost Platform for City-Scale Sensing}

\author{Joshua Adkins}
\affiliation{
  \institution{University of California, Berkeley}
  %\city{Berkeley}
  %\state{California}
}

\author{Branden Ghena}
\affiliation{
  \institution{University of California, Berkeley}
  %\city{Berkeley}
  %\state{California}
}

\author{Neal Jackson}
\affiliation{
  \institution{University of California, Berkeley}
  %\city{Berkeley}
  %\state{California}
}

\author{Pat Pannuto}
\affiliation{
  \institution{University of California, Berkeley}
  %\city{Berkeley}
  %\state{California}
}

\author{Samuel Rohrer}
\affiliation{
  \institution{University of Michigan}
  %\city{Ann Arbor}
  %\state{Michigan}
}

\author{Bradford Campbell}
\affiliation{
  \institution{University of Virginia}
  %\city{Charlottesville}
  %\state{Virginia}
}

\author{Prabal Dutta}
\affiliation{
  \institution{University of California, Berkeley}
  %\city{Berkeley}
  %\state{California}
}

\renewcommand{\shortauthors}{Adkins, Ghena, Jackson, Pannuto, Rohrer, Campbell, and Dutta}

\begin{abstract}

City-scale sensing holds the promise of enabling a deeper understanding of our
urban environments.
However, a city-scale deployment requires physical installation,
power management, and communications---all challenging tasks standing between a
good idea and a realized one.
This indicates the need for a platform that enables easy deployment and
experimentation for applications operating at city scale.
To address these challenges, we present \name, a modular, energy-harvesting platform for city-scale sensing.
\name simplifies deployment by eliminating the need
for connection to wired infrastructure and instead harvesting energy from an
integrated solar panel. The platform furnishes the key resources necessary to 
support multiple, pluggable sensor modules while providing
fair, safe, and reliable sharing in the face of dynamic energy constraints.
We deploy \name with several sensor modules, showing the viability
of an energy-harvesting, multi-tenant, sensing system, and evaluate its
ability to support sensing applications.
We believe \name reduces the difficulty inherent in city-scale deployments,
enables new experimentation, and provides improved insights into urban health.

\blfootnote{For questions, email \href{mailto:adkins@berkeley.edu}{adkins@berkeley.edu}}

\end{abstract}

\maketitle

\section{Introduction}
\label{sec:intro}
\placefigure[t]{fig:intro:signpost}

\begin{dfigure}{fig:intro:signpost}
  \centering
  \includegraphics[width=\columnwidth]{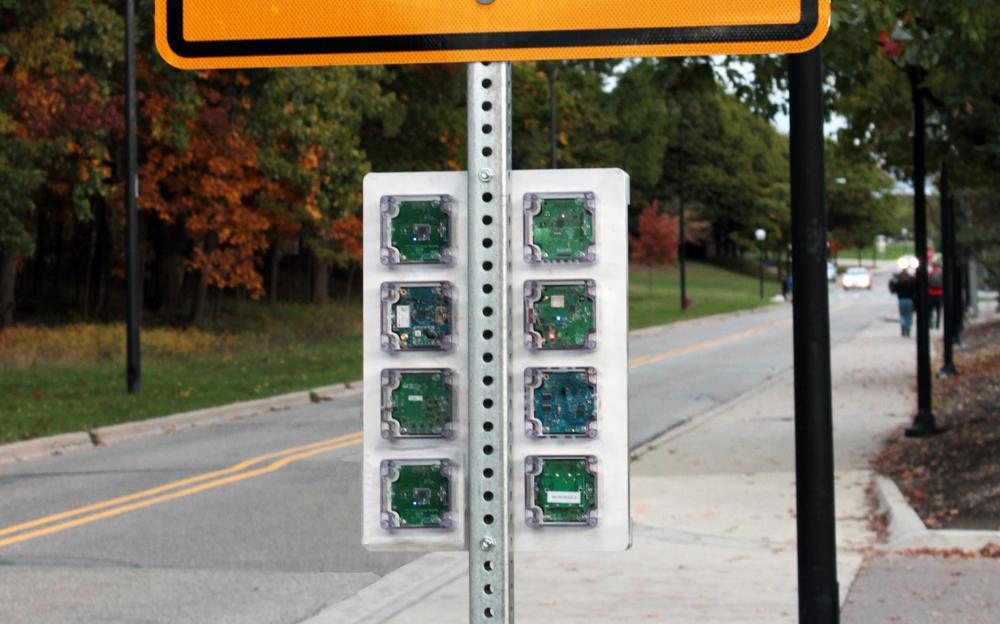}
  \caption{%A deployed \name.
    \normalfont
    The \name platform easily mounts to existing street sign posts, harvests
    from an integrated 0.1\,\mm solar panel, and provides tenant sensor modules
    with power, communications, processing, storage, time, and location.
    %with power, networking, storage, time, location, and higher performance compute.
    \name is open source, with all hardware and software available
    online.$^\dagger$
  }
\end{dfigure}

Today, more than 50\% of the world's population live in urban areas, and the U.N.\
projects that to increase to 66\% by 2050~\cite{un14pop}. With increasing population
density, there is growing interest in making cities safer, cleaner,
healthier, more sustainable, more responsive, and more efficient---in a word,
smarter. Supporting this interest are numerous funding
opportunities~\cite{nsfsmartcities,eusmartcities,chinasmartcity}, interested
cities~\cite{smartcolumbus,denversmartcity,li2015development}, and active research
projects~\cite{mydlarz2017implementation,catlett2017array,cheng2014aircloud,ledeczi2005},
all targeting new technology to enable smarter cities. And for good reason:
applications such as pedestrian route planning based on air quality, noise
pollution monitoring, and automatic emergency response alerts can all improve
the quality of life for a city's inhabitants.

However, we believe that the difficulty of deploying existing smart city
technology and applications is impeding progress.
Deployments are rooted in single-purpose hardware, necessitating redesigns to
support upgraded sensors or revised goals.
Moreover, each system requires a re-implementation of standard resources such
as power, communications, and storage, taking developer time away from the core
application.
Deploying sensors is difficult too, with the reliance on energy from
wired mains constraining installation locations.
These problems limit not only production-ready technology, but also make it
particularly challenging to perform short-term, exploratory research, speaking
to the need for a platform that will lower the barrier to entry.

\blfootnote{$\dagger$ \url{https://github.com/lab11/signpost}}

To address these challenges, we present \name, a modular, energy-harvesting
platform enabling deployable city-scale sensing applications.
It mounts to pervasive sign posts (\cref{fig:intro:signpost}) and harvests
energy from a vertically mounted solar panel.
To reduce the burden of developing new applications, \name provides commonly
required services including power, communications, processing, storage, time, and
location.
The platform is modular, with eight pluggable slots for sensors,
processors, and radios, facilitating modifications and upgrades to the system.
To enable shared deployments, \name is multi-tenant, supporting multiple
applications simultaneously and enforcing isolation between them.

Key to \name's deployability is its energy-harvesting, modular
architecture.
Harvesting energy enables the system to sever ties to wired infrastructure.
This in turn opens up an increased selection of deployment locations, allowing
for more granular deployments. Harvesting also enables short-term, pop-up
deployments to drive application development and experimentation.
Support for modularity allows the sensors on the platform to be changed to suit
application needs. More fundamentally, however, modularity permits \name to take
advantage of future technology improvements, improving its capabilities over
time.

An energy-harvesting, multi-tenant platform faces challenges that do not exist
for mains-powered, single-purpose systems.
For one, eliminating the connection to mains power limits the energy
available for sensing. We assess the expected solar energy throughout
the US,
finding that a module can expect an average power of at least 120--210\,mW for 50\% of weeks.
We also provide APIs allowing software to adapt to existing energy, reducing
functionality in times of famine and opportunistically increasing it when
possible.
Another challenge is managing and sharing platform resources to support
multiple stakeholders with unaligned interests. We explore the hardware and
software requirements for measuring usage and enforcing isolation, describing
guarantees necessary for sharing \name's limited energy budget between
applications.

We envision a testbed of \names supporting short-term experimentation by many
users.
\names have been deployed on the University of California, Berkeley
campus for six months. The ongoing deployment monitors weather, senses
TV whitespace spectrum usage, and observes vehicular traffic.
We have found \name modules are generally easy to create and the software API
is simple to implement on commonly used software and hardware platforms such as
Arduino and ARM Mbed.
To facilitate the creation of new hardware and software to run on \name, we have
also created desktop development kits capable of emulating deployed behavior.
We hope that by providing a platform for city-scale sensing that reduces the
barriers to deployable applications, supporting that platform with development
tools and accessible interfaces, and working with the community to realize
their sensing needs, we can gain deeper insight into the workings of urban
areas and enable higher-level applications that impact policy and quality of
life throughout a city.

         % 1.0 pg
\section{Related Work}
\label{sec:related}

Existing work in urban sensing generally falls into three categories:
static deployments of sensing applications,
mobile or human-based participatory sensing,
and---most similarly to \name---%
deployments of generic sensing infrastructure.
The first two categories are particularly insightful as a guide to which
services are frequently needed by existing applications, which we summarize in
\cref{tab:related:services}.

Examples of static deployments include
acoustic sensors to monitor, characterize, and
localize different sounds~\cite{mydlarz2017implementation, girod2006, ledeczi2005},
particulate sensors to monitor air quality~\cite{cheng2014aircloud}, and
electromagnetic, radiological, and meteorological sensing to track
people~\cite{li2015senseflow} and cars~\cite{abari2015caraoke}, measure road
conditions~\cite{bouillet2013fusing,sen2012kyun}, monitor wireless
traffic~\cite{rose2010mapping}, locate point sources of
radiation~\cite{rao2008identification}, and identify severe weather in urban
environments~\cite{basara2009overview}.
Most deployments are not long-term and are only
deployed for the purposes of evaluation.
Additionally, almost all of these deployments depend on either mains power or a battery
for an energy source, motivated by the desire for rapid prototyping.
Many of these deployments use a proof-of-concept node design built
mostly with off-the-shelf components, without much consideration for optimized
energy consumption~\cite{girod2006,mydlarz2017implementation,sen2012kyun,
ledeczi2005, li2015senseflow, bouillet2013fusing}.
By providing a platform that already handles energy-harvesting,
\name could provide sustainability to these deployments.
Further, based on reported power numbers, with the
exception of the high power Micronet nodes
\cite{basara2009overview,illston2009design}, many of these applications and
experiments could run on the \name platform without significant
redesign.

\placefigure[t]{tab:related:services}
\begin{dtable}{tab:related:services}
    \centering
    \begin{adjustbox}{width=\columnwidth}
        \begin{tabular}{ l | c !{\color{lightgray}\vrule} c !{\color{lightgray}\vrule} c !{\color{lightgray}\vrule} c !{\color{lightgray}\vrule} c !{\color{lightgray}\vrule} c !{\color{lightgray}\vrule} c !{\color{black}\vrule}}
        \textbf{Deployment} & \textbf{Energy}  & \textbf{Network} & \textbf{Processing} & \textbf{Storage} & \textbf{Time} & \textbf{Sync} &  \textbf{Location} \\ \hline
        Caraoke \cite{abari2015caraoke} & \textbf{\checkmark} & \textbf{\checkmark} & & & \textbf{\checkmark} & & \\ \arrayrulecolor{lightgray}\hline
        Bouillet et al. \cite{bouillet2013fusing} & \textbf{\checkmark} & \textbf{\checkmark} & & & & & \\ \arrayrulecolor{lightgray}\hline
        AirCloud \cite{cheng2014aircloud} & \textbf{\checkmark} & \textbf{\checkmark} & & & & & \\ \arrayrulecolor{lightgray}\hline
        Girod et al. \cite{girod2006} & \textbf{\checkmark} & \textbf{\checkmark} & \textbf{\checkmark} & & & \textbf{\checkmark} & \textbf{\checkmark} \\ \arrayrulecolor{lightgray}\hline
        L\'edeczi et al. \cite{ledeczi2005} & \textbf{\checkmark} & \textbf{\checkmark} & \textbf{\checkmark} & & & \textbf{\checkmark} & \textbf{\checkmark} \\ \arrayrulecolor{lightgray}\hline
        SenseFlow \cite{li2015senseflow} & \textbf{\checkmark} & \textbf{\checkmark} & & & & & \\ \arrayrulecolor{lightgray}\hline
        Argos \cite{rose2010mapping} & \textbf{\checkmark} & \textbf{\checkmark} & & & \textbf{\checkmark} & & \\ \arrayrulecolor{lightgray}\hline
        SONYC \cite{mydlarz2017implementation} & \textbf{\checkmark} & \textbf{\checkmark} & \textbf{\checkmark} & \textbf{\checkmark} & & & \\ \arrayrulecolor{lightgray}\hline
        Kyun Queue \cite{sen2012kyun} & \textbf{\checkmark} & \textbf{\checkmark} & & \textbf{\checkmark} & \textbf{\checkmark} & & \\ \arrayrulecolor{lightgray}\hline
        Micronet \cite{illston2009design} & \textbf{\checkmark} & \textbf{\checkmark} & & \textbf{\checkmark} & & & \\ \arrayrulecolor{lightgray}\hline
        Seaglass \cite{ney2017seaglass} & \textbf{\checkmark} & \textbf{\checkmark} & & \textbf{\checkmark} & \textbf{\checkmark} & & \textbf{\checkmark} \\ \arrayrulecolor{black}\hline

    \end{tabular}
    \end{adjustbox}
    \caption{Services required by existing applications.
        \normalfont
        %By providing these services, \name can support existing applications
        %and simplify the creation of new ones.
        Time is millisecond-accurate as
        provided by services like NTP, while Sync is microsecond-accurate as
        provided by GPS. Location is GPS-level accurate coordinates.
        These represent the minimum services a platform should provide to
        support existing applications and simplify the creation of new ones.
        % corresponds applications that expect GPS-level accurate location
        % estimates.
        %We note that applications that do not require time or
        %highly precise location services could still benefit from leveraging
        %these resources to reduce the burden of manually labeling deployment
        %locations and the uncertaincy and jitter introduced by timestamping
        %data upstream of its collection.
        %
        Many of these applications could run on \name without significant
        modifications.
    }
\end{dtable}

The majority of work that targets urban sensing uses
participatory methods, in which users participate
with mobile phones and other handheld devices~\cite{burke2006participatory,
campbell2006people}, or vehicles are outfitted with various
sensors~\cite{lee2006efficient, hull2006cartel}. These methods use
existing mobile resources to collect similar data to static deployments.
Many have paired mobile phones with
handheld air quality monitors~\cite{bales2012citisense,dutta2009common,devarakonda2013real,cheng2014aircloud},
or used phones to directly meter noise pollution~\cite{rana2010ear, maisonneuve2009noisetube,
bilandzic2008laermometer} or traffic conditions~\cite{mohan2008nericell,thiagarajan2009vtrack,work1710mobile}.
Similar to participatory sensing methods, vehicular sensor networks
monitor air quality, traffic, and road
conditions~\cite{lee2006mobeyes, hull2006cartel,mathur2010parknet,eriksson2008pothole,devarakonda2013real},
and even detect rogue cellular base stations~\cite{ney2017seaglass}.
These types of deployments often scale very well as the mobility of the devices
allows a few sensors to reach a much larger area.
However, incentivizing participation can be difficult and coverage can
be unpredictable and potentially insufficient.

Finally, several platforms provide generic sensing infrastructure,
suitable for many types of smart city applications.
CitySense proposes an open, city-scale
wireless networking and sensor testbed~\cite{murty2008citysense}. It utilizes
mains-powered, street pole mounted embedded Linux nodes with 802.11 mesh
networking and enables in-situ node programming by end users. Argos, a
passive wireless mapping application, builds on a 26 node CitySense
deployment~\cite{rose2010mapping}. Unfortunately, the CitySense architecture
met many logistical challenges that ultimately limited a scaled
deployment~\cite{mattconvo}. The Array of Things project utilizes a
network of sensor nodes distributed throughout Chicago to gather environmental
data including light, temperature, humidity, and air quality~\cite{catlett2017array}. Like CitySense, Array of Things sensor
nodes assume wired power and networking, and thus must be installed in
locations where these resources are present.
\name also provides an open testbed for smart city research.
However, through its focus on deployability and modularity, \name
reaches a different design point than these projects, resulting in a
resource-constrained, energy-harvesting, and multi-tenant platform
that is more easily deployed, but potentially more challenging to program.

       % 1.0 pg
\section{Platform Overview}

\begin{dfigure}{fig:overview:overview}
  \centering
  \includegraphics[width=\columnwidth]{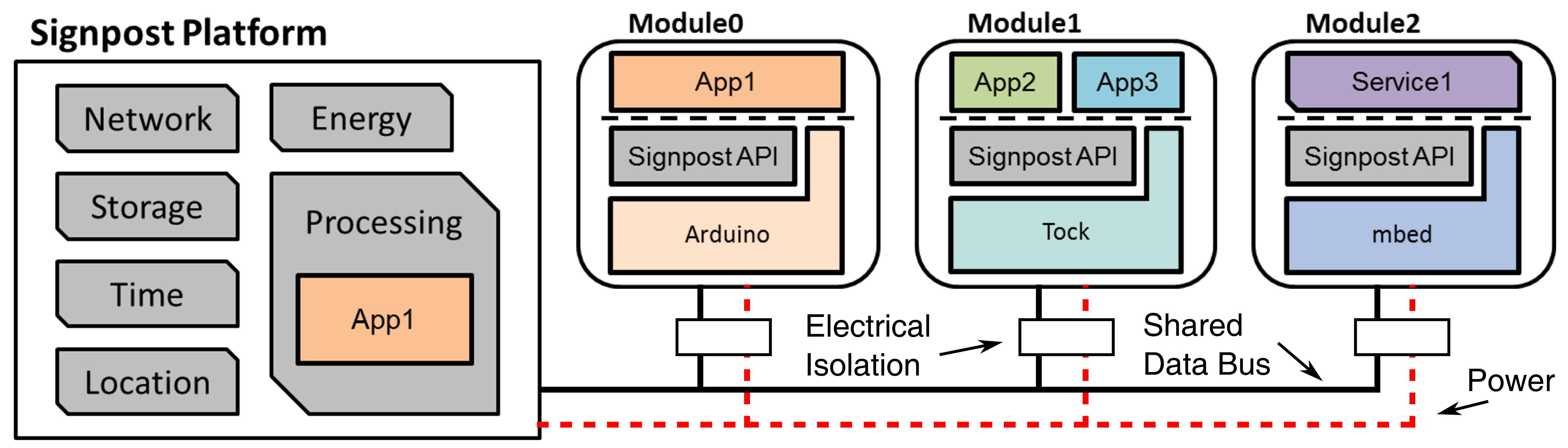}
  \caption{\name platform overview.
    \normalfont
    \name monitors and distributes energy to connected modules and provides
    shared networking, Linux processing, storage, time, and
    location services. Modules implement one or more sensing modalities and utilize many
    possible software stacks, running one or more applications or even
    providing additional services to the platform. Applications can potentially
    be distributed across the platform and modules. This platform design supports
    development and deployment of urban sensing applications.
    }
\end{dfigure}
\placefigure[t]{fig:overview:overview}

\label{sec:goals:overview}

In the following sections, we present the design, implementation, and
evaluation of \name, a modular, solar energy-harvesting, sensing platform.
In the \name platform, sensor hardware connects to a
shared
backplane
via a standard electrical and mechanical interface, enabling modularity.
The backplane
serves as the module interconnect
and has the ability to
electrically isolate each module, allowing energy use of any particular
module to be limited.
To support these sensor modules,
the platform harvests solar energy, monitors a shared battery, and distributes
metered power.
It provides multiple radio interfaces
for different communication patterns and shares them
among the modules. Other services are implemented as well,
including time and location, data storage, and compute offload using a
Linux-class co-processor, and these services can be accessed by modules
through a standard software API. Resources and modules are orchestrated by a
microcontroller-based system controller that oversees the operation of the
\name platform.  All of these components are housed in a waterproof aluminum
case that to bolts to a standard street sign post for easy deployment.
\Cref{fig:overview:overview} shows an overview of the platform.

      % 0.5 pg
\section{Design}
\label{sec:design}
The \name platform's design is guided by four high-level goals:

\noindent
\begin{itemize}[leftmargin=*]
  \item \textbf{Deployability}
    is the primary concern of the platform and is key
    to enabling larger and more frequent deployments, and ultimately
    wider adoption by the community.

  \item \textbf{Accessibility}
    reduces burden for developers, thus the platform needs to provide services
    that meet common application needs.

  \item \textbf{Modularity}
    allows developers to modify and extend sensing capabilities to
    support new applications and upgrade modules as technology improves.

  \item \textbf{Multi-tenancy}
    enables the platform to simultaneously host mutually-untrusting
    applications
    %and modules
    created by multiple stakeholders, reducing
    deployment burden and the cost of experimentation.
    %
    %The ability to co-locate applications and sensors created by multiple
    %stakeholders further reduces deployment burden.
    %
    %The co-location of applications and sensors created by parties with
    %potentially unaligned interested further reduces deployment burden.
    %
    %The platform needs to support applications and sensors created by parties
    %with potentially unaligned interests.

\end{itemize}

\subsection{Deployability}
\label{sec:design:deploy}

Deployability is the primary concern for the \name platform.
Many urban sensing applications require fine-grained sensing, which is not
possible for platforms that can only be deployed with easy access to mains
power or wired networking.
Additionally, to support ad-hoc experimentation, the platform needs to be easily
installed, removed, and moved. A deployment made today may not meet the sensing
needs of an application tomorrow.

In order to enable deployability, \name does not depend on mains power or wired
networks. Relying on wired infrastructure would limit \name deployments to
locations with grid access, such as the top of streetlight poles, and would
require costly and time-consuming installation by city utility workers.
To support easy physical installation, the platform attaches
to existing infrastructure found ubiquitously in urban areas---sign posts.

Making these deployability decisions allows \name to better support some applications
while restricting others, particularly applications with high power sensors,
significant bandwidth needs, or heavy computation. To address these concerns,
the platform needs to provide software primitives that enable applications to
adapt to available energy and bandwidth. Even if these primitives prove
insufficient, we believe that in time most applications will still
become possible on \name due to the rapid power scaling of embedded hardware.
In the last decade alone, best-in-class microcontroller active current has decreased from
220\,\uA/MHz to 10\,\uA/MHz~\cite{msp,apollo}, radio transmission power has reduced by 3-5x~\cite{atrf,nrf}, and
many sensors have followed similar trajectories.
By embracing modularity, hardware can be updated to capitalize on these improvements,
with the tradeoff between deployability and resource constraints increasingly
favoring the \name architecture.

\subsection{Accessibility}
Informed by a review of prior sensing projects in \cref{sec:related}, \name
provides several services to support accessibility and reduce the
burden for application developers.

\subsubsection{Energy}
Since wired mains power is not an option for \name, we turn to batteries and energy
harvesting to power the system. Batteries alone may be sufficient for
short-term research deployments, but replacement is not scalable for
geographically distributed deployments. Instead, a battery would need to store
enough energy for the entire deployment duration. Assuming a 1\,cm thick Li-ion
battery the size of the \name solar panel (0.096\,m$^2$) yields a storage
capacity of 576\,Wh~\cite{jeong2011prospective}.
For one year of lifetime, this would result in an
average platform power budget of 66\,mW.

The expected budget can be improved significantly with the addition of solar
energy harvesting. An optimally oriented, 17\% efficient solar panel with the
same area as \name's would generate 2.4\,W on average indefinitely in Seattle, a
city with notably poor solar conditions~\cite{nrelsolar}.
Even with vertical panel placement and sub-optimal panel orientation, the
addition of energy harvesting yields an increase in energy provided to the
platform as we demonstrate in \cref{sec:eval:energy:harvesting}, resulting in increased application capabilities.

\subsubsection{Communications}
\name needs to support periodic data transmissions, firmware updates, and
occasional bulk data uploads. Coverage
is needed over a wide area and neither wired network nor WiFi access points can
be expected to be accessible for all deployed \names.
One solution to these problems is cellular radios, especially the
machine-to-machine focused LTE Cat-1, LTE-M, or NB-IoT networks. Cellular
networks provide high throughput and good coverage, but also come with costs, both in
terms of high power draw and network usage fees.

Alternative solutions include low-power, wide-area networks such as
LoRaWAN~\cite{lora}, which provides data transfer at rates of 1-20\,kbps with a
range of several kilometers and power draw significantly lower than cellular
radios. LoRaWAN networks can be deployed by end users, allowing a network to be
set up to support a \name deployment. However, LoRaWAN predominately supports
uplink communications, making firmware updates and other downlink-focused
applications more difficult.

Finally, local communication
facilitates interactions
between a \name and any nearby residents
or users of the platform. Communication protocols such as Bluetooth Low Energy would
enable the platform to interact directly with nearby smartphones.

\subsubsection{Processing}

In nearly any sensing system, data must be processed, batched, transformed, and
analyzed, and in the face of energy constraints, local
computation is preferable over transferring all data to the cloud.
Providing a processing service is not necessarily just about computational
capability. A familiar processing environment in which developers can
use familiar languages and libraries lowers the barrier to entry
for domain scientists.

Many existing urban sensing platforms provide processing by using some
variation of a Linux computer as their primary
processor~\cite{murty2008citysense, rose2010mapping, catlett2017array,
mydlarz2017implementation, rao2008identification}. For an energy-constrained
system, however, supporting an always-on Linux computer is problematic. Even
the lowest power Linux compute modules we survey draw 200-500\,mW when
active~\cite{IntelEdison}.
One compromise is to use a Linux environment not as a
core controller, but as a co-processor, employed occasionally to process
batched data. This allows developers to use languages and libraries to which they are
accustomed, but requires them to split applications between two execution
environments.

\subsubsection{Storage}
With low power and low cost flash memory widely available, data storage could
be a module-supplied resource. However, we argue it should be centralized on \name for
two reasons.
First, a central data store aids manual data collection (likely over a
short-range wireless link). This is useful for collecting high-fidelity
data from multiple modules, particularly in the early experimentation phases of
a deployment.
Second, co-locating the central storage with shared processing resources
allows for fast and easy access to batched data.

\subsubsection{Time and Location}
Synchronizing clocks throughout a sensor network deployment is critical to many
applications~\cite{sundararaman2005clock}. Providing the capability to
synchronize within 100\,ns allows a group of \names to achieve
localization within 30\,m for RF signals and less than one meter for audio
signals.
In addition to just synchronization, the ability to timestamp data and
understand the local time of day and year is useful for adapting operation
(for example, slowing sampling before night) or predicting available solar
harvesting energy.
Location also provides automatic installation metadata and enables
localization-based applications, such as gunshot detection.
Fortunately, all are easily provided by GPS modules, although some care needs to
be taken when expecting GPS use in dense city environments where fewer
satellites may be in line-of-sight of the receiver.
The addition of a stable and low power real-time clock can act as an optimization for a time and
location system on a stationary platform by allowing the GPS to be predominantly
disabled. This reduces system power draw while maintaining sufficient
accuracy for many applications.

\subsection{Modularity}
Modularity enables not only specialization, but
it also allows the platform
to be upgraded over time, adapting to technology improvements for sensor modules and
platform resources alike.
Supporting modularity requires standardized electrical and mechanical
interfaces to allow sensor modules to be installed and replaced as
needed. The electrical interface should be simple but sufficient, including
connections to power and an internal communication bus over which modules
access platform services.
Other signals can be added to support performance, for example a
time synchronization signal, but such additions should be kept to a minimum to keep
module creation simple.

Regarding mechanical considerations, the interface must allow for a robust connection to the
physical platform without significantly limiting sensing capability. Weatherproofing
plays an important part in the design of this interface since \name will be
deployed outdoors, as does physical security since platforms will be unattended
for long periods. Additionally, sensor module developers should be able to
easily tailor the module enclosure to support the physical and
environmental requirements of their sensors.

\subsection{Multi-tenancy}
Finally, \name is designed to support multiple stakeholders simultaneously,
allowing a single hardware deployment to act as a testbed for multiple applications.
Support for multi-tenancy requires fair sharing of resources between applications.
For most system services, this reduces to platform software recording
usage and implementing some fairness policy.

Sharing energy is a more complex problem and the top priority of a
multi-tenant, energy-harvesting system~\cite{adkins2017energy}. The power
requirements of one application should not limit the capabilities of another.
To support this, a platform must first be able to accurately measure and
control access to energy. This involves metering not just modules, but also
system resources, so that their energy draw may be charged against the
application which accessed them.

Second, the platform must use these measurements to implement an energy policy.
In the presence of variability, applications need guarantees of energy
availability to reason about future processing capabilities.
There is one important guarantee: the energy allocated to an application must
only decrease in a predictable fashion. It can be spent directly by the
application, indirectly by a service the application uses, or taken regularly
as a platform tax, but it must not decrease in a manner unpredictable to the
application. Particularly, energy should never be taken to support other
applications (although it could be given). If energy is harvested by the
platform, the allocation of a particular application may increase, but having a
minimum known energy to rely on allows applications to plan for future actions.
Support for energy isolation has been explored in prior
work~\cite{adkins2017energy}.

Features to support multi-tenancy have an added benefit in supporting overall
system reliability. Modules can be isolated from the platform entirely if a
hardware or software failure occurs.
        % 3.0 pg

\section{Implementation}
\label{sec:impl}

\begin{dfigure}{fig:impl:architecture}
    \centering
    \includegraphics[width=\columnwidth]{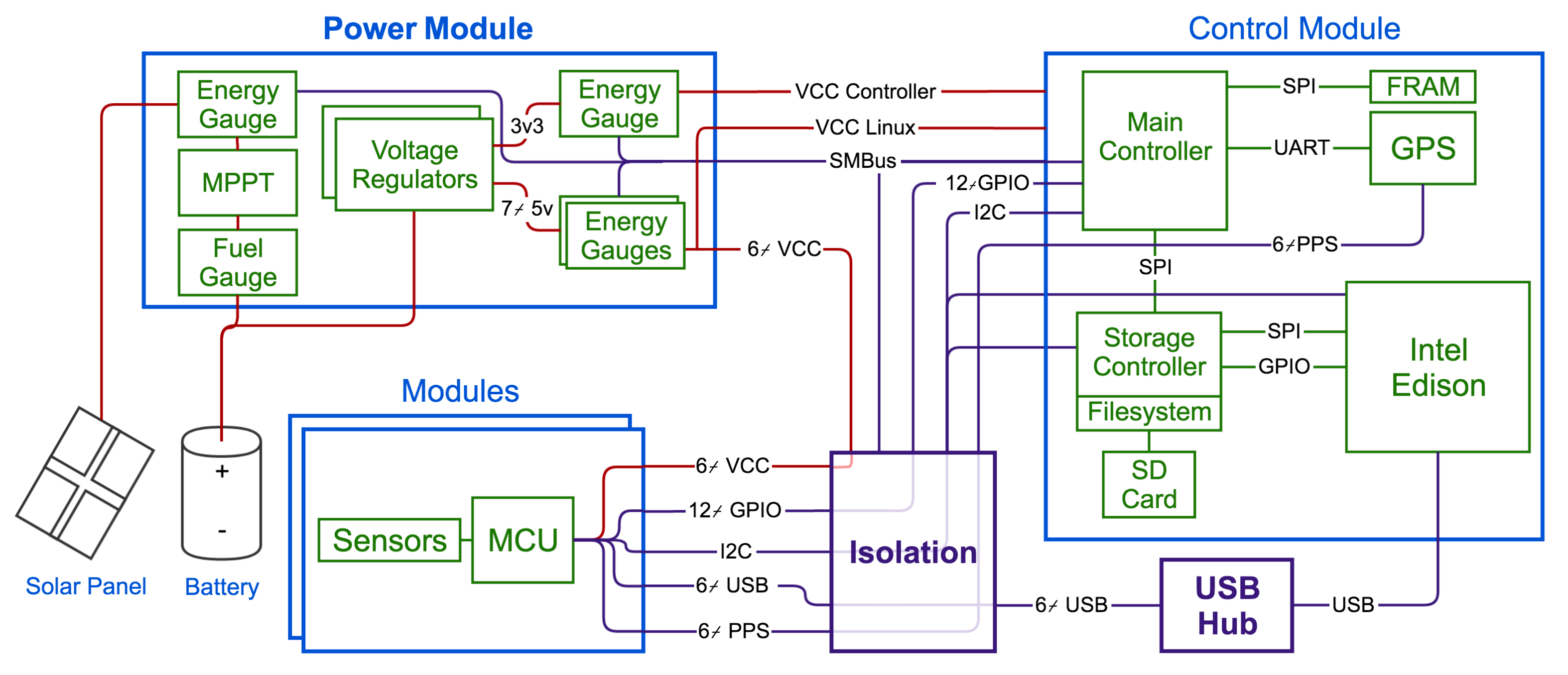}
    \caption{\name architecture.
        \normalfont
        %Each \name has three parts. First, a
        The Power Module is
        capable of harvesting energy from a solar panel, storing energy in a
        battery, supplying power at the correct voltage to modules, and
        monitoring the energy use of modules.
        %Second, a
        The Control Module
        provides storage, time and location, and Linux processing services, and also
        monitors modules with the capability of isolating them from the system
        if necessary. Finally, there are the modules themselves, with many
        possible capabilities. This architecture allows for modular and extensible sensing
        while minimizing deployment complexity.
    }
\end{dfigure}

The \name architecture is shown in \cref{fig:impl:architecture}. The \name
platform is defined by the Power Module, Control Module, Backplane, and Radio
Module.
Additional modules connect via a standard electrical and mechanical interface.
A full \name has six general-purpose module slots, one of which is taken by the Radio
Module, leaving five for sensing capabilities. The size of the entire system,
including a case, is 42.9\,cm high, 30.0\,cm wide, and 8.4\,cm thick.
For comparison, the minimum size of a
speed limit sign in the United States is 91\,cm by 61\,cm~\cite{adminstration2009manual}.

\subsection{Backplane}

The Backplane is the backbone of the \name. It has physical and electrical
connections for modules, signal routing between modules, and isolation
hardware.
The Backplane has eight slots in which modules can be connected. Two are
special-purpose, corresponding to dedicated signals for the Power Module and
Control Module. The remaining six are standard interfaces for modules. The
interface provides power at 5\,V, access to a shared \iic bus, two dedicated
I/O lines to the Control Module, a Pulse Per Second (PPS) signal for synchronization,
and a USB slave connection.

Modules are not required to implement all signals in this interface.
However, we expect that most modules will
use the \iic bus and dedicated I/O signals, and that some complex modules will
implement USB or PPS support.

All module connections can be individually isolated, along
with buffering for \iic connections. These isolators
can be activated by the Control Module and prevent individual modules
from negatively impacting the rest of the \name.
The Backplane also
accepts a voltage reference signal from each module and handles translation of
voltage levels for all signals except USB, allowing modules to perform I/O at
any voltage between 1.65\,V and 5\,V.

\subsection{Power Module}

The Power Module is responsible for energy harvesting, management, monitoring, and
distribution on the \name platform. Energy is harvested from a Voltaic
Systems 17 W solar panel, a 37\,cm by 26\,cm panel with an expected 17\%
efficiency.
The solar panel output is monitored by a coulomb counter, and regulated
by a maximum power point tracking battery charger. Excess energy is stored
in a custom 100\,Wh Li-ion battery pack.

System energy is further regulated for consumption before being distributed to
the Backplane and modules.
Each regulator can provide a constant 1.5\,A, and is protected from shorts
by a load switch. Each module's power rail is monitored by a coulomb counter
that also provides instantaneous current readings,
supporting energy accounting.

\placefigure[t]{fig:impl:architecture}

The Power Module also
includes a hardware watchdog that monitors the platform.
This further increases \name
reliability by providing a redundant watchdog in
the event of software failures.

\subsection{Control Module}

The Control Module handles system tasks, such as managing the module energy
usage, assigning module addresses, and monitoring system faults.
It also provides time, location, storage, and processing services to the
sensor modules.
Computation is handled by two ARM Cortex-M4 microcontrollers.

One microcontroller is responsible for isolation, managing
the GPS, and accounting for module energy. It can also communicate with sensor modules
on the shared \iic bus and through dedicated per-module I/O signals, sending
information such as location and time to the sensor modules in response
to \name API calls. A globally synchronized Pulse Per Second signal is
routed from the GPS to all sensor modules.
The second microcontroller is responsible for managing an SD card and providing the
storage API to the sensor modules.
Each of these subsystems is power gated and can be entirely disabled to save energy.

Finally, the Control Module has an Intel Edison Linux compute module for higher
performance processing capabilities.
Contrary to common system design, while the Edison is the most capable computer
on the \name, it is not in control of the system.
Instead, the Edison is a coprocessor, capable
of batch processing and using languages and libraries that are difficult or impractical
to port to embedded microcontrollers.
The Intel Edison connects directly to modules over USB, with
each module playing the role of a USB slave device. It can also communicate
with modules over an internal SPI bus by using one of the Cortex-M4s on the Control Module
to forward messages to the shared \iic bus.
The power usage of the Edison is individually monitored, allowing its energy
to be attributed to the module utilizing its services.

\subsection{Radio Module}

The Radio Module provides communications services to the \name. To
handle diverse communication needs, it hosts
cellular, LoRa, and BLE radios. An ARM Cortex-M4 microcontroller handles receiving
messages through the shared \iic bus or via USB from the Intel Edison and
sending them to the appropriate radio interface.
A U-blox SARA-U260 cellular radio
is capable of both 2G and 3G operation at up to 7.2\,Mb/s.
However, it
draws up to 2.5\,W in its highest throughput modes~\cite{ubloxu2}.
A Multitech xDot radio module provides LoRaWAN communications.
Sending data through LoRaWAN is more sustainable from an energy budget
standpoint, with the module drawing less than 0.5\,W in its highest power
state~\cite{xdot}.
Finally, the Radio Module includes an nRF51822 BLE SoC.
This enables \name to send real-time data about the
environment to nearby smartphones.
Providing three communications interfaces allows \name to make decisions about
which radio to use based on quality of service, latency, throughput, and
energy requirements.

\placefigure[t]{fig:impl:development_backplane}
\subsection{Sensor Modules}
Four sensor modules have been created for \name and are in use. The
existing modules perform ambient environmental sensing
(temperature, humidity, pressure, and light), monitor energy in seven audio frequency bins ranging from 63\,Hz to 16\,kHz, measure RF
spectrum usage within 15\,MHz to 2.7\,GHz, and detect motion within 20\,m with a microwave radar. Each was made by a
different student, including two undergraduates.
All of the sensor modules and the \name Backplane
are shown in \cref{fig:impl:development_backplane}.

\placefigure[t]{tab:impl:apis}
\subsection{Module Software}

\begin{dtable}{tab:impl:apis}
    \centering
    \begin{adjustbox}{width=\columnwidth}
    \begin{tabular}{lll}
\hline
\textbf{Service}    & \multicolumn{1}{l}{\textbf{System Call}}                 & \multicolumn{1}{l}{\textbf{Description}} \\ \hline
\textbf{Init}       & \texttt{i2c\_address = module\_init(api\_handles)}       & Initialize module                        \\ \hline
\textbf{Network}    & \texttt{response = network\_post(url, request)}          & HTTP POST data to URL                    \\
                    & \texttt{network\_advertise(buf, len)}                    & Advertise data over BLE                  \\
                    & \texttt{network\_send\_bytes(destination, buf, len)}     & Send via best available medium           \\ \hline
\textbf{Storage}    & \texttt{record = storage\_write(buf, len)}               & Store data                               \\ \hline
\textbf{Energy}     & \texttt{energy\_info = energy\_query( )}                 & Request module energy use                \\
\textbf{}           & \texttt{energy\_set\_warning(threshold, callback)}       & Receive energy usage warning             \\
\textbf{}           & \texttt{energy\_set\_duty\_cycle(duty\_cycle)}           & Request duty cycling of module           \\ \hline
\textbf{Processing} & \texttt{processing\_call\_rpc(path, buf, len, callback)} & Run code on Linux compute                \\ \hline
\textbf{Messaging}  & \texttt{messaging\_subscribe(callback)}                  & Receive message from a module            \\
                    & \texttt{messaging\_send(module\_id, buf, len)}           & Send message to another module           \\ \hline
\textbf{Time}       & \texttt{time\_info = get\_time( )}                       & Request current time and date            \\
\textbf{}           & \texttt{time\_info = get\_time\_of\_next\_pps( )}        & Request time at next PPS edge            \\ \hline
\textbf{Location}   & \texttt{location\_info = get\_location( )}               & Request location                         \\ \hline
    \end{tabular}
    \end{adjustbox}
    \caption{\name API examples.
      \normalfont
      Abstract versions of several \name API calls for each system service
      are shown.
      Providing a high-level API enables easier application
      development.
    }
\end{dtable}

To enable access to the resources on \name, we provide APIs for
applications that abstract away the specific details of messages sent over
the shared \iic bus and allow module creators to write software at a higher
level. Abstract versions of several API calls are listed in
\cref{tab:impl:apis},
including calls to allow module applications to POST data, write to an
append-only log, be automatically duty-cycled, start
processes on the Intel Edison, and send messages to other modules.

All API calls are layered on a minimal intra-\name network protocol.
The library code is written in C on top of a hardware
abstraction layer requiring \iic master, \iic slave, and GPIO implementations.
We implement the library using the Tock operating system~\cite{tockos} for
our own development purposes and have ported the library to the Arduino~\cite{arduino}
and ARM Mbed~\cite{mbed} stacks to support a wider array of module designs.

\name supports multiple views on what it means to be an application.
A module may run one or more applications,
and an application may be constrained to a single module, include processing
code run on the Intel Edison, exist logically across several modules
connected by the messaging API, or even across \names distributed around a
city, connected by wireless communications.  An example of the \name software
model is shown in \cref{fig:overview:overview} where one or more applications
are running on heterogeneous sensor modules and accessing \name services
through a common API.

\subsection{Development}

\begin{dfigure}{fig:impl:development_backplane}
    \centering
    \begin{subfigure}{0.45\columnwidth}
        \centering
        \includegraphics[width=\textwidth]{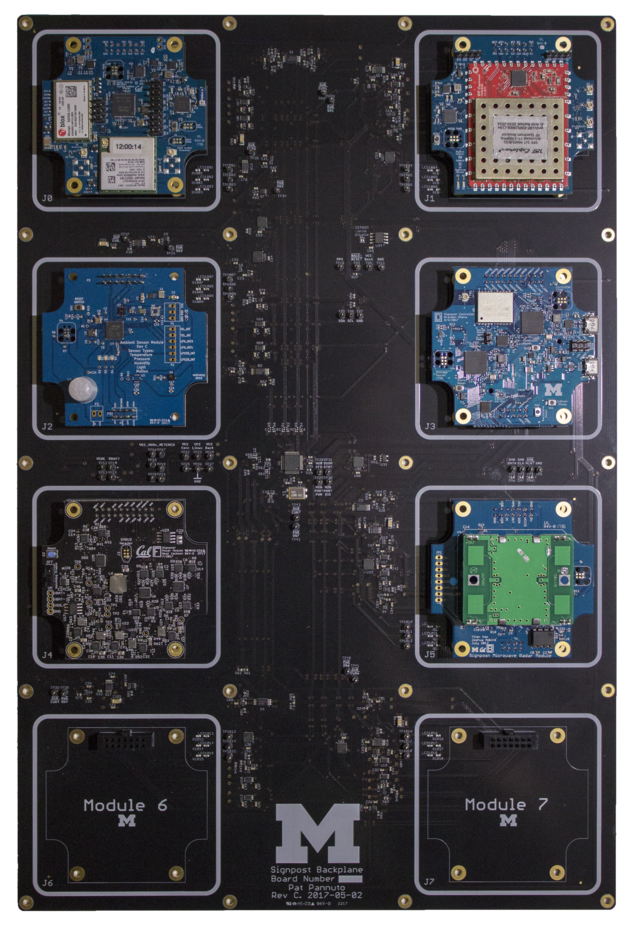}
        \caption{}
    \end{subfigure}
    \begin{subfigure}{0.45\columnwidth}
        \centering
        \begin{subfigure}[b]{\textwidth}
            \centering
            \includegraphics[width=0.53\textwidth]{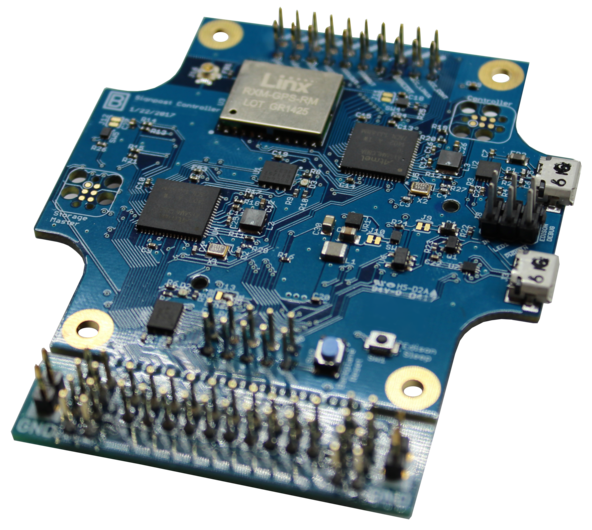}
            \caption{}
        \end{subfigure}\\
        \begin{subfigure}{\textwidth}
            \centering
            \includegraphics[width=0.8\textwidth]{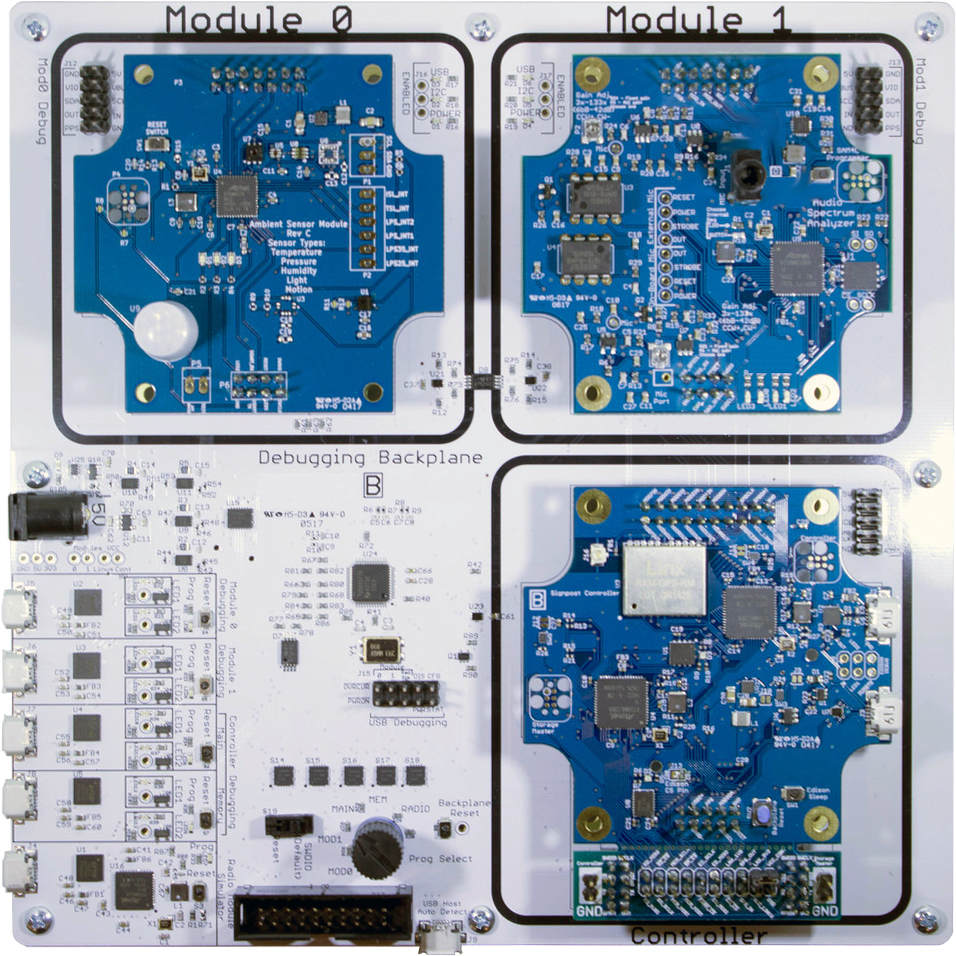}
            \caption{}
        \end{subfigure}
    \end{subfigure}
    \caption{A populated Backplane (a), Control Module (b) and Development Backplane (c).
        \normalfont The Backplane serves as the \name interconnect, while
        the smaller Development Backplane is the desktop equivalent,
        enabling easy module and application creation and testing. The Control
        Module manages \name energy and provides services to sensor modules. Existing sensor
        modules are also shown, with the RF spectrum and radar modules at the top and bottom
        right of the populated Backplane respectively, and the environmental and audio sensing
        modules on the top left and top right of the Development Backplane.
    }
\end{dfigure}

\placefigure[t]{fig:eval:solar_harvesting}

In addition to the full, weatherproofed \name platform,
a
development version of the system aids in creating and testing module
hardware and software, as shown in \cref{fig:impl:development_backplane}. The
development \name supports two modules and a Control Module.
While meant to be wall-powered, it has the same isolation and monitoring
hardware as a full \name, allowing it to emulate various energy states,
track module energy use, and disable modules when they exceed their
allocation.
Rather than including radios, the development \name has a microcontroller that
implements the radio API, but sends data over a USB serial connection
instead of an RF link.
Identical Control Module and Backplane hardware is used on both systems,
allowing desktop experimentation with applications that is faithful to
deployed system.

          % 2.5 pg
\section{Evaluation}
\label{sec:eval}
We evaluate the key claims of the \name platform, including
deployability, the implications of a deployable design on energy availability,
and the ability to support multiple applications. We also benchmark
several \name services.
Finally, given these capabilities and constraints, we explore the types of
applications capable of running on \name and how they interact with system
resources.

\subsection{Deployment Metrics}
A primary goal of the \name platform is deployability, and over the course
of nine months we deploy the platform on over 50 occasions, for varying
lengths of time, at several locations. In all of these deployments,
we found \name to meet our deployability goals in both speed and effort.

Specifically, we find that two students can deploy a single \name in less
than five minutes. In a specific case, it took less than 90 minutes to walk
and deploy twelve \names across a portion of the UC Berkeley campus.
Although we take no precautions, the deployments have
experienced no vandalism or theft, even with
\names placed near a popular concert venue in an area with relatively
high property crime.
We believe that this indicates the platform is unobtrusive and blends in with
other city infrastructure. Approval for these deployments, while sometimes
slow for bureaucratic reasons, has been simple due to the non-destructive,
attachment method. While this level of deployability comes at the cost of energy
availability, a system with these properties greatly facilitates
ad-hoc experiments and highly-granular long
term sensing applications.

\subsection{\name Energy}
This focus on deployability makes energy availability a fundamental challenge for \name. We
investigate the overhead of multi-tenancy and expectations
for how much energy \name can harvest.

\placefigure[t]{fig:eval:solar_ccdf}
\subsubsection{Platform Overhead}
\label{sec:eval:energy:overhead}

While supporting city-scale sensing is the purpose of \name, not all energy
goes directly to applications. In particular,
multi-tenancy and platform services each incur overhead.
These costs can be primarily attributed to the static power of the
regulation and monitoring hardware, which have a total quiescent power
draw of 13.2\,mW. The components for module isolation
draw an additional 1\,mW, as do the microcontrollers on the Control Module, on average.

Additionally, the over-sized charging and regulation circuitry
has a lower efficiency than similar circuitry designed
to match the requirements of a single-purpose sensor. We measure
the battery charging efficiency to be 85\% at a wide range of power inputs,
and the regulator efficiency to be 89\% at all but the lowest power draws.
Across the platform, this totals to 76\% efficiency and a base power draw of 16\,mW, less
than 2\% of the 50th percentile average power budget and 6-18\% of
the 95th percentile budget. We believe this is an acceptable
overhead for the advantages of multi-tenancy.

Services provided by the Control Module, such as storage and location,
are power gated when not in use and do not contribute to the static power
of the platform.
If applications request these services, their energy is attributed to the
sensor modules using them.
We find the Intel Edison Linux module draws 15--24\,mW in sleep mode, the GPS
chip draws 40\,mW when tracking satellites,
and the Radio Module sleeps at less than 1\,mW. The SD card
is enabled on demand, and therefore has no idle power draw.

\subsubsection{Harvesting}
\label{sec:eval:energy:harvesting}

A key enabler of deployability is the shift to a solar energy-harvesting
power source. To further increase deployability, it is preferable to make no assumptions about
solar panel positioning, and therefore expect the panel to be deployed vertically
facing an arbitrary direction. We evaluate the expected energy availability
under these constraints in different locations, solar panel directions,
and times of year.

We start this evaluation by deploying four solar panels on sign posts in Ann Arbor, Michigan,
with one panel pointing in each cardinal direction. A building is located to the south of the
posts and a small hill directly west. For each panel, we record the open-circuit voltage and
short-circuit current at ten second intervals and estimate the power output of
the panels by assuming an 80\% fill factor. \Cref{fig:eval:solar_harvesting} shows
the output of this experiment for one week in July 2016 and one week in March 2017. We present both the
instantaneous output of each solar panel and the daily averages.

This experiment shows that the power availability of a \name is highly variable,
ranging from over 3.08\,W for the south facing panel on March, 22nd to only
219\,mW for the north facing panel on March, 25th. We find
that the direction, season, and degree of cloud cover all contribute to this variability.
While some of the variability can be buffered by the battery, variability
will inevitably be experienced by applications running on \name.

To more broadly determine the expected power budget for a sensing application
running on \name, we create an energy availability model that predicts
the average weekly power available to \names at different geographic
locations in the United States throughout the year.
The model is based on hourly direct and diffuse
light measurements at 1,500 locations around the United States from the
NREL MTS2 dataset~\cite{nrelmts2}, and these
measurements are converted into expected power output using a standard
harvesting model for tilted solar panels which takes into account solar
panel direction, angle, and the harvestable portion of diffuse light~\cite{solarmodel}.
We compare our model with the experimental data shown in \cref{fig:eval:solar_harvesting}
and find the model strictly underestimates our experimental results by an average 3.3\%
on sunny days and 22\% on cloudy days. We believe this error
primarily can be attributed to diffuse light collection for north-facing
solar panels, a scenario that is not well studied in solar modeling literature.

The results of this model are displayed in \cref{fig:eval:solar_ccdf} as
the fraction of weeks at which an application
will have a minimum available power. To generate this plot, we group the weekly
average power data by latitude, subtract the platform overhead and regulator
efficiency losses discussed in \cref{sec:eval:energy:overhead}, then divide
by an expected five applications (assuming one for each available module slot on \name).
In addition to showing data for each latitude,
we also plot energy available in Seattle, Washington and San Diego, California,
which are particularly
poor and ideal solar energy harvesting locations, respectively, in the United States.
We see that the \nth{95} percentile of available weekly average power ranges from
3.84\,mW per application for a north facing \name in Seattle, WA to 147\,mW per
application for a south facing \name in San Diego, CA.

\placefigure[t]{fig:eval:energy_distribution_trace}
We conclude that, in general, the direction at which \name is placed impacts
available energy more than the latitude of the platform. This creates a
tradeoff between deployability and energy availability.
While it is possible to entirely ignore
orientation when deploying \names, this comes at the cost of expected energy for some of
the deployed systems.
Putting in care to avoid facing north when possible may be a sufficient
compromise.

One aspect which is not included in the prior evaluations is potential shading
from nearby obstructions. This is a particularly real concern in urban areas
where buildings are expected to obstruct direct sunlight for portions of each
day. The amount of shade a \name can expect is, however, particularly
deployment-specific and difficult to predict in a general fashion. For example,
due to its vertical orientation, even with a building directly to its east a
west facing \name can expect to harvest most of its predicted clear-sky
energy. In our deployments, we have found that \names deployed under moderate,
continuous shade (under a tree in this case) see harvested energy similar to a
north facing, clear-sky \name.

\begin{dfigure*}{fig:eval:solar_harvesting}
    \centering
    \begin{subfigure}{0.73\textwidth}
	\centering
	\includegraphics[width=\linewidth]{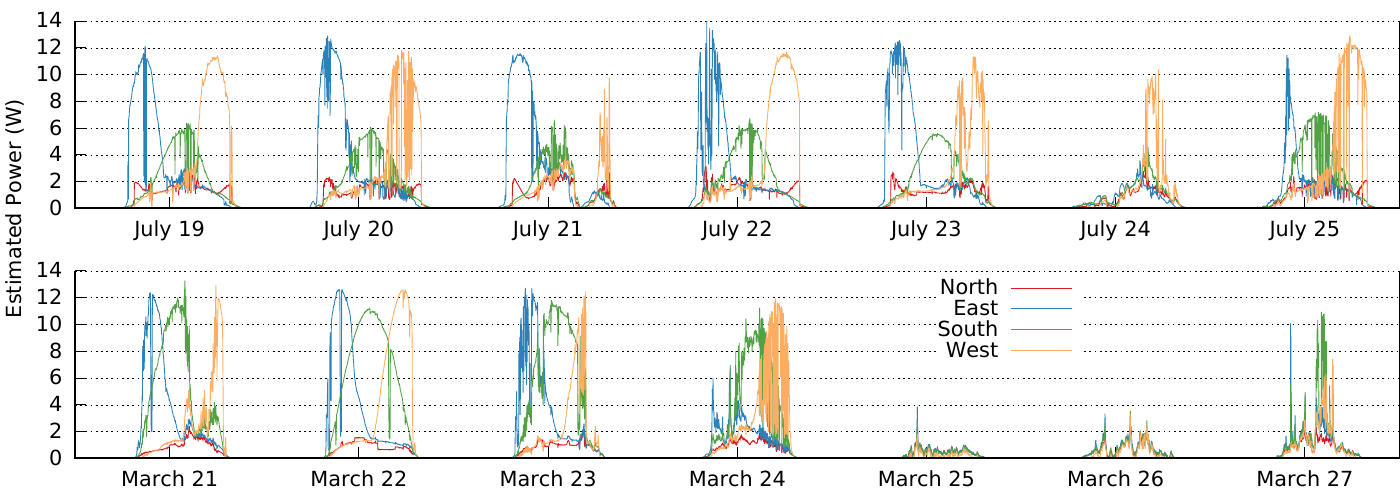}
    \end{subfigure}
    \begin{subfigure}{0.26\textwidth}
	\centering
	\includegraphics[width=\linewidth]{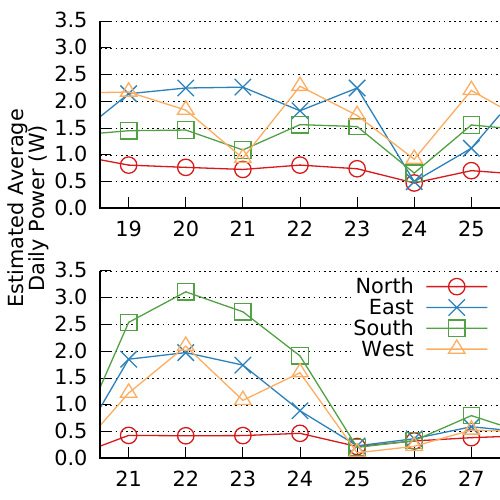}
    \end{subfigure}
    \caption{Solar harvesting in four different cardinal directions and two seasons.
	\normalfont
	The experiments are run in July 2016 and March 2017 in Ann Arbor, Michigan,
    with each including periods of both sunny and
	cloudy days.
	At left is estimated power generated from solar panels mounted
	vertically in four cardinal directions captured in 10 second intervals
	over a week.
	At right is the average daily power provided by each solar panel.
    There are large variations in average power both due to direction and
    daily weather patterns. While some daily variations can be buffered by the
    battery, \name will still experience variability in available energy to which
    it must adapt.
    }
\end{dfigure*}

\begin{dfigure*}{fig:eval:solar_ccdf}
  \centering
  \includegraphics[width=\textwidth]{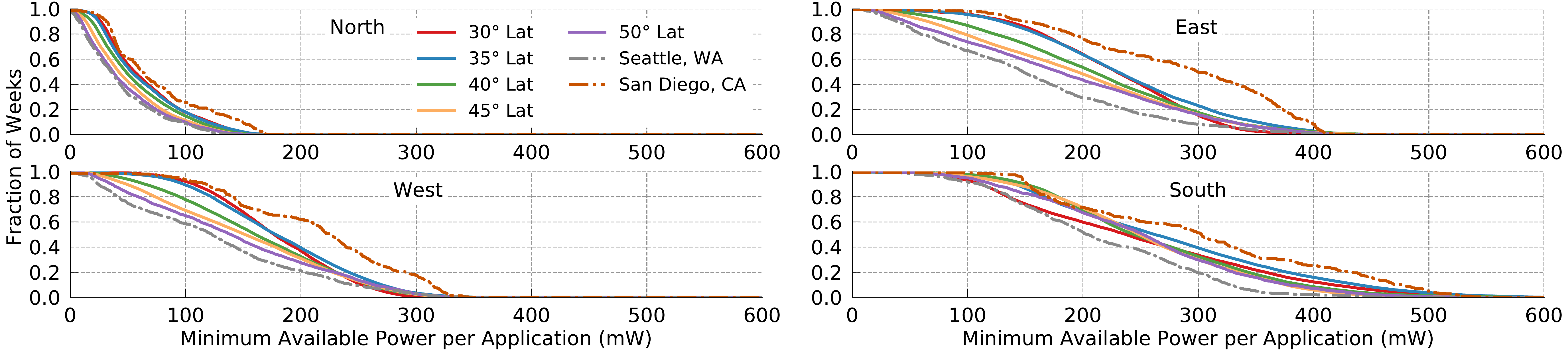}
  \caption{Fraction of weeks when an application can expect a minimum power income
    at different latitudes and cardinal directions.
    \normalfont
    To evaluate how much power a \name application can expect under
    varying deployment conditions, we model the solar harvesting
    %To better understand the distribution of power that \name applications may
    %receive under varying deployment conditions, we model the solar harvesting
    potential of a vertical \name facing the four cardinal directions across
    the United States.
    We use a standard solar model that accounts for both direct and diffuse
    light~\cite{solarmodel} along with hourly irradiance data from the NREL
    MTS2 2005 dataset~\cite{nrelmts2}.
    We group these locations by latitude, and also plot distributions for
    Seattle, Washington and San Diego, California, where local weather patterns
    create poor and near-ideal solar harvesting conditions, respectively.
    %
    %The plots show the fraction of weeks where
    %each module will receive at least a specific amount of power.
    The per application expected minimum power is calculated by subtracting the
    static power draw (16\,mW) from the weekly average harvested power,
    dividing among an expected five applications, and
    multiplying by the regulator efficiency (76\%).
    %
    %We find
    %the daily average power of the model agrees with our \hl{experimental data within 7\% for similar times, locations
    %and weather conditions}.
    We find that orientation generally has a stronger influence on harvested
    energy than latitude or climate.
    }
\end{dfigure*}

\subsection{Managing Multi-tenancy}
\name expects to host not just a single application, but several. Here, we
evaluate how the system responds to multiple demands to its resources
simultaneously.

\subsubsection{Energy Isolation}

\begin{dfigure*}{fig:eval:energy_distribution_trace}
    \centering
    \includegraphics[width=\textwidth]{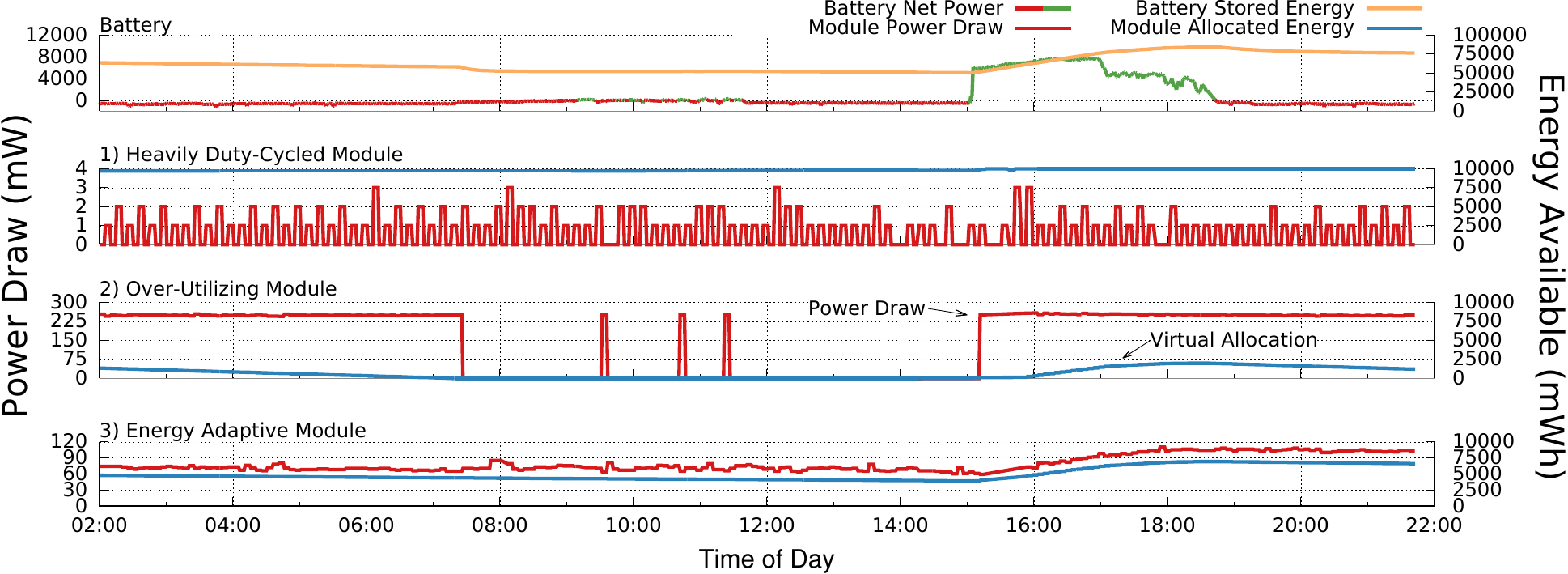}
    \caption{Energy isolation on \name.
      \normalfont
      Energy allocation and five-minute average power draw are displayed for
      three simultaneously running applications and the platform as a whole.
      Each application employs a different strategy for energy use. The first
      is only active for a brief period every ten minutes, achieving a low
      average power, and storing up an allocation of energy. The second
      continuously runs, exhausting its budget, and is disabled by the
      platform, to be enabled later when energy is available again. The third
      adapts its actions based on the available energy, running continuously
      without depleting its allocation.
      \name is capable of balancing the needs of these three applications
      simultaneously, assigning each a ``virtual allocation'' of energy it
      draws from without affecting the operation of the others.
    }
\end{dfigure*}

The primary resource that must be shared between all applications is energy.
On \name, we virtualize stored energy, making it appear to each application
that they have independent batteries. Stored energy in the battery is split
into a ``virtual allocation'' for each application.
A virtual allocation is guaranteed to never deplete except when predictably
spent. For
example, it will never be taken to support another application's needs.
This allows programs to plan and make decisions based on available energy that
are independent of the actions and needs of others.

On an energy-harvesting platform, an additional question arises in how to
distribute incoming energy.
A fair model distributes energy equally between applications, but there must be
a maximum allocation for each. If an application stores the energy it is given
but does not use it, its allocation would eventually expand to the entire
capacity of the battery.
Instead, we define a maximum capacity for each virtual allocation. Harvested
energy is then divided between applications that are below maximum capacity.
This adds variability to the amount of energy an application receives based on
the actions of other applications running on the platform. However, this
variability is no worse than the variability inherent to energy harvesting
systems in the first place.
Policy choices and support for energy isolation are discussed further
in another work~\cite{adkins2017energy}.

\Cref{fig:eval:energy_distribution_trace} demonstrates energy sharing in
practice. Three modules are installed on one \name, each running a single
application and given virtual allocations with a maximum capacity of
10,000\,mWh. Data is shown for a 20 hour period, from night to night. The
deployed \name has a building directly to the east, only allowing it to harvest
later in the day.
Displayed are the five-minute average power draws for each application and the
net power into the battery. Energy allocations are also reported every five
minutes for each module and the battery.

Each application has a different strategy for energy use.
The first heavily duty-cycles itself and is active for only a brief period
every ten minutes. This results in an average power draw of less than 4\,mW,
and consequently its virtual allocation stays near or at maximum capacity the
entire time.
The second application continuously draws 250\,mW, an amount that cannot be
sustained while the \name is receiving no direct sunlight. It eventually
exhausts its allocation and is disabled by the platform. Later in the day,
when energy is being harvested, it is allocated a portion of incoming energy
and resumes operation.
The third application adapts to the amount of energy available to it, remaining
in continuous operation. Its power draw increases when the solar panel receives
direct light, corresponding to an increase in sampling rate in the application.
As this experiment demonstrates, \name is able to isolate the energy needs of
applications from each other.

\subsubsection{Internal Communication}

The \name design includes a single, shared, multi-master \iic network for
internal communication, such as requests to platform services. When multiple
applications are running simultaneously, this bus can be a source of
contention.
While the \name design expects only a modest utilization of the shared \iic
bus, in practice sensing events can often be correlated and traffic can be
bursty.
Theoretically the listen-before-talk requirement of \iic should make the bus
achieve nearly 100\% reception rates even in these scenarios, however we
observe that this feature is not implemented in all TWI/\iic peripherals.
Assuming no carrier sense capability, the \iic bus resembles the original
unslotted ALOHAnet~\cite{abramson1970aloha}, and the target utilization rate
should be kept to the 20\% proposed by ALOHA. This corresponds to a total
traffic of 80\,kbps on a 400\,kHz \iic bus, which we believe is sufficient for
most sensing applications.
Applications that require higher throughput can make use of the optional USB bus.

\subsection{Microbenchmarks}

Several services are important to benchmark due to their
impact on the range and performance of \name applications.

\subsubsection{Communication Policy}

\begin{dfigure}{fig:eval:comm_policy}
  \centering
  \includegraphics[width=\columnwidth]{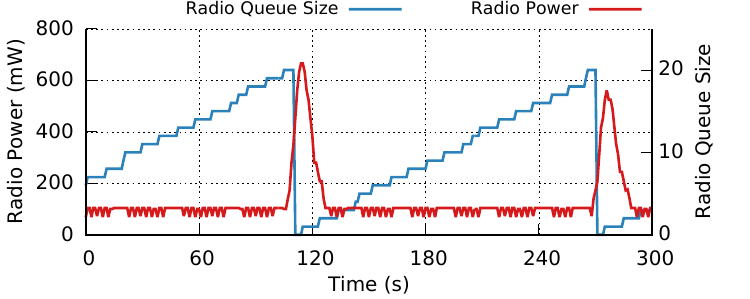}
  \caption{Communication policy in practice.
    \normalfont
    The power draw of the Radio Module is shown along with the number of
    messages queued to be sent. The communication policy is set to
    automatically transfer data over a cellular connection if the
    queue reaches twenty messages, as can be seen by the increased power draw.
    This policy allows the platform to adapt to both increased application
    requests and poor network conditions by utilizing high-power resources.
  }
\end{dfigure}
\placefigure[t]{fig:eval:comm_policy}
\placefigure[t]{fig:eval:app_resources}

\name provides multiple wireless interfaces. These have an advantage in supporting
various communication policies that determine how data should be be transmitted
based on quality of service needs and the current energy state of the platform.
One simple policy is to primarily use the lower power LoRaWAN radio for data
transmission unless the message queue gets too full, which could occur when
applications have large amounts of data to transfer or in poor radio conditions
when LoRaWAN bandwidth is limited.
When the queue gets too full, the cellular radio is activated and all queued
messages are transferred quickly.
In \cref{fig:eval:comm_policy}, we demonstrate an example of this policy. Poor
communication conditions are emulated by removing the LoRaWAN radio antenna,
causing messages to be queued until the cellular radio is activated to dispatch
them, resulting in briefly increased power draw.

\subsubsection{Synchronization}

Some applications require coordination between
multiple modules on a single \name or between multiple \names, requiring tight
synchronization~\cite{sundararaman2005clock}.
On \name, a PPS signal is routed to each of the sensor
modules from the GPS to provide this synchronization.
We find the timing difference across \names to be 75\,ns in the
average case with a \nth{95}\,percentile metric of 97\,ns.
We observe little skew in the signal from Control Module to sensor modules (less
than 6\,ns) and almost no variation from module to module.
We expect this synchronization precision to suffice for many applications,
providing sufficient resolution for RF localization on the order of tens of
meters and sub-meter audio localization.

\subsection{Applications}

\begin{dfigure}{fig:eval:app_resources}
  \centering
  \includegraphics[width=\columnwidth]{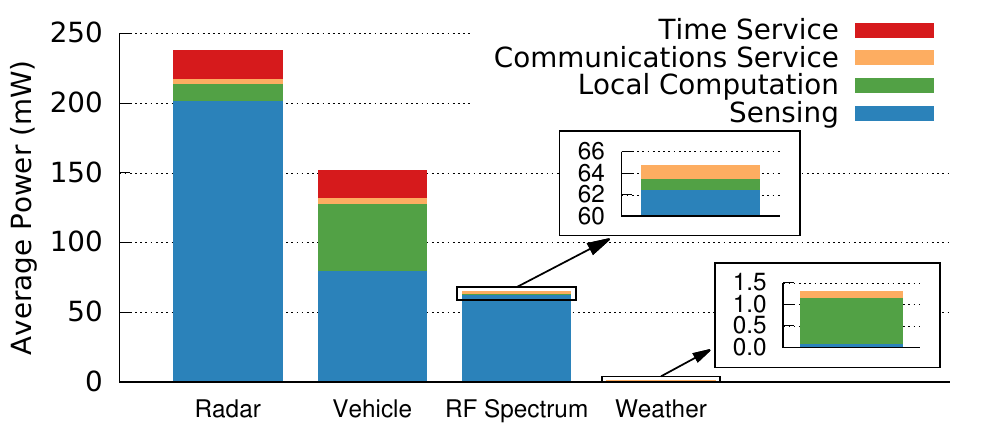}
  \caption{Resource usage of example applications.
    \normalfont
    We break apart the major components of usage for example applications into
    sensing cost, local computation, and network and time service requests.
    %The components that make up this total are the cost of the sensor, local
    %computation on the module, usage of the networking service, and usage of
    %the time service.
    %
    Heavily duty-cycled applications such as the weather monitoring app have
    nearly inconsequential average power.
    Applications performing constant sensing with tight timing requirements
    both draw a higher total power and remit a greater share platform power draw.
    %such as time (which is split between the radar and vehicle counting
    %applications in this example).
    Applications like spectrum sensing can achieve moderate average power draw
    even with high instantaneous sensing power using duty cycling.
    Dynamically adjusting duty cycling allows spectrum sensing to adapt
    to energy availability.
  }
\end{dfigure}

Applications run on sensor modules and have access to system resources through
physical connections and software APIs.
We design several applications (and sensor modules) and deploy them on
the Berkeley campus for several months.
While applications written by users will be different, these examples can
inform the types of applications that are possible on \name. We describe our
applications, the platform resources they use, and some example results.
\cref{fig:eval:app_resources} shows the power drawn by different components of
the applications, broken down into draw by sensors, local processors, and
the communications and time services.

\subsubsection{Weather Monitoring}
The weather monitoring application uses the environmental
sensing module to sample temperature, pressure, and humidity
every ten minutes, sending it to the cloud via the \name network API.
After the data reaches the cloud, it is posted to Weather Underground to help
support their goal of distributed weather sensing. The application achieves
very low power operation even without implementing sleep mode by using the
energy API to power off the sensor module between samples.

\begin{figure}
  %\FrameSep5pt
  %\begin{framed}
  \begin{tcolorbox}[colframe=white,colback=offwhite,boxrule=0.5pt,arc=4pt,left=6pt,right=6pt,top=6pt,bottom=6pt,boxsep=0pt,width=\columnwidth]
\begin{scriptsize}
\lstset{language=C,
  basicstyle=\ttfamily,
  keywordstyle=\color{blue},
}

\lstset{
  emph={uint8_t,api_t, NULL, time_t},
  emphstyle={\color{blue}},
  commentstyle={\color{asparagus}},
}

\begin{lstlisting}[aboveskip=-0.5\baselineskip,belowskip=-0.5\baselineskip,language=C]
uint8_t send_buf[DATA_SIZE];

void send_samples (void) {
  // Add a timestamp to the data
  time_t time = get_time();
  memcpy(send_buf, time, sizeof(time_t));

  // Send data over network, allowing Signpost to decide how
  network_send_bytes(send_buf, DATA_SIZE);
}

int main (void) {
  // Initialize the module with Signpost
  api_t* handles[] = NULL; // provides no services
  module_init(handles);

  // Collect audio data with an ADC, placing it into send_buf
  adc_continuous_sample(SAMPLE_RATE, &data_ready_callback);

  // Send samples every ten seconds
  timer_every(10000, &send_samples);
}
\end{lstlisting}
\end{scriptsize}
  \end{tcolorbox}
  %\end{framed}
  \caption{Example module software.
    \normalfont
    This software snippet from the vehicular sensing application collects
    averaged volume data for ten seconds and transmits it using the
    network API. Timestamps for the collected data are requested from
    the time API and appended to the data before transmitting it.
    Access to the \name APIs makes applications easier to create.
  }
  \label{fig:eval:code_example}
\end{figure}

\subsubsection{Vehicle Counting}
The vehicle counting application runs on the audio sensing module,
which provides the volume of audio in seven frequency bins collected up to
100 times per second. This module should in principle allow high-level event recognition
(e.g. vehicle detection), without capturing recognizable human speech.
The application records these volumes, averages them over a second,
and every transmits the results every ten seconds to the cloud using the
network API. To properly
identify vehicle movement, the application must know the precise time at which
a volume sample is taken, so the time API is used to timestamp
each batch.
The code for this application is shown in \cref{fig:eval:code_example},
and the average power draw and resource usage are shown in \cref{fig:eval:app_resources}.
The requirement for precise timing information results in the application being charged
for a portion of the GPS power. Additionally, the local processor must stay active with
a relatively high clock frequency to continually sample and process incoming audio volume
data.
Once the data are in the cloud, it is further processed to look for peaks
that are indicative of a moving car. An example of the output of this processing is
shown in \cref{fig:eval:vehicle}.

\placefigure[t]{fig:eval:vehicle}
\placefigure[t]{fig:eval:rf-spectrum}

\subsubsection{RF Spectrum Sensing}
The white space sensing application runs on the RF spectrum module and
periodically samples the energy on each of the TV white space channels (every 6\,MHz
from 470-830\,MHz). For thirty seconds, the spectrum analyzer reads the energy on
these channels and computes the min, max, mean, and standard deviation for
each channel. The application then sends this data with the \name network
API and uses the energy API to power off.  While the duration for power
off is currently set to three minutes, it could be adapted to available energy
without significantly degrading the utility of the application.

Three days of this data are shown for several interesting channels in
\cref{fig:eval:rf-spectrum}. While our RF spectrum module does not yet meet
the FCC requirements for a white space utilization sensor, collecting distributed
RF spectrum data can be used to inform RF propagation models and inform policy
about the reuse of underutilized spectrum.

\begin{dfigure}{fig:eval:rf-spectrum}
  \centering
  \includegraphics[width=\columnwidth]{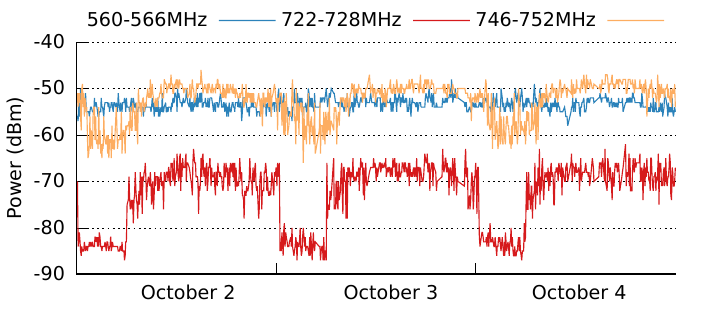}
  \caption{RF spectrum sensing application.
    \normalfont
    A sample of RF spectrum data from October 2017 in three frequency bands corresponding to a
    local TV station (560\,MHz), AT\&T owned spectrum (722\,MHz), and Verizon
    owned spectrum (746\,MHz). Distributed
    and fined-grained spectrum sensing could help to build better models
    of RF propagation and inform policy around the reuse of underutilized
    spectrum. The two higher frequency bands are particularly interesting
    due to their cyclic nature.
  }
\end{dfigure}

\begin{dfigure}{fig:eval:vehicle}
  \centering
  \includegraphics[width=\columnwidth]{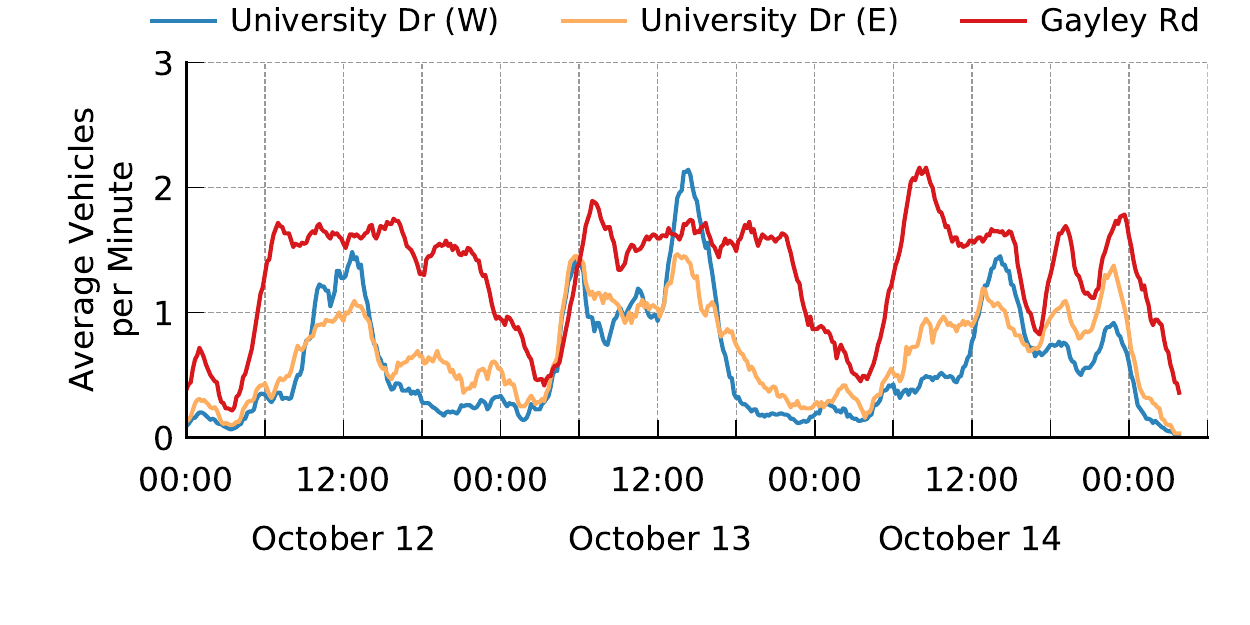}
  \caption{Vehicle counting application.
    \normalfont
    Several days of processed audio data are
    collected in October 2017 for the vehicle counting application.
    %Audio volume data is collected and filtered, then prominent peaks across several
    Prominent peaks across several audio
    frequency bands are used to detect vehicles.
    We plot estimated vehicles per minute averaged over a one hour time window.
    The \names on University Drive are close, but do not
    %The \names on the smaller campus road are quite close to one another, but do not
    have completely redundant traffic paths.
    %This explains their similar, but
    %not exact match in estimated traffic.
    We note that Gayley Road sees traffic much later into the night
    because it is a through street that routes around campus. Interestingly,
    all the \names experience traffic until around midnight on October 14th,
    and after further examination, this was due to a concert at a nearby
    venue. Clear peaks in traffic can be seen before and after the concert,
    which started at 20:00.
  }
\end{dfigure}

          % 3.0 pg
\section{Discussion}
\label{sec:disc}

\name is under active development.
Here we discuss on-going issues
along with future work for the platform.

\subsection{Cost}
\label{sec:disc:cost}

Currently \name costs roughly \$2,000 to produce in quantities of ten, including
all parts and labor for a \name platform and a typical set of
sensor modules.
We expect the price to drop significantly
at higher quantities, and we plan to explore optimizations to further reduce
cost.
For context,
including labor, a street sign and post costs \$250, a solar powered electric
speed limit sign costs \$3,000, yearly maintenance costs for
a stop light are \$8,000, and a new stop light costs over \$250,000~\cite{streetsign,solsign,stoplight}.
This puts \name on par with other city infrastructure.

\subsection{Community Building}
\label{sec:disc:developer}

To realize the benefits of \name modularity,
domain experts must be motivated to leverage the platform.
To achieve developer buy-in, we believe
we need to create a suite of tools that help people develop and test
both hardware and software at their desk, then allow them to deploy it on
existing \names. We have started
with the Development
Backplane described in \cref{sec:impl},
and we plan to continue to grow the ecosystem around it.
Additionally, we have ported the \name software API to
platforms such as Arduino and Mbed which are more accessible
to non-experts.
All of our hardware and software is
open source to encourage the creation of a community around the
\name platform. Software, hardware and documentation for \name
can be found at \href{https://github.com/lab11/signpost}{github.com/lab11/signpost}.

\subsection{Security and Privacy}
\label{sec:disc:privacy}

The pervasive deployment of sensors throughout a city creates significant
privacy concerns. While our sensors cannot collect personally
identifiable information, the platform could enable the collection of this data
if care is not taken. This problem must be addressed through both policy and practice. 
Policy should dictate
that modules are incapable of collecting private information and that applications
do not attempt to collect or transmit sensitive data. In practice, this
requires some manual oversight into the module creation and deployment process and 
that all software updates to \name be authenticated to ensure they
originate from the proper source.
Additionally, the data collected from \names must be authenticated to ensure its validity,
especially if it will directly influence city infrastructure.
However, the authentication of large numbers of low-power sensors in a collaborative
deployment is an area of active research.

          % 0.25 pg
\section{Conclusions}
\label{sec:conc}

In this paper, we introduce \name, a solar energy-harvesting modular platform designed to
enable city-scale deployments.
By providing energy, communications, storage, processing, time, and location
services, \name
allows developers to focus on the sensing application they care about
rather than the engineering details of making it deployable.
The platform is designed with adaptivity in mind, giving applications the tools
to adjust to varying energy and communications availability.

By making the \name platform widely available, we hope to begin a new era of
urban sensing. We envision a future where city-scale experimentation
is simple and city-scale deployments are pervasive.
This in turn will open new areas of research
exploring energy constrained, geographically distributed applications,
encouraging the development of more capable sensors,
and providing a deeper understanding of our increasingly urban world.

          % 0.25 pg
\section{Acknowledgments}
\label{sec:ack}

This work was supported in part by the CONIX Research Center,
one of six centers in JUMP, a Semiconductor Research Corporation
(SRC) program sponsored by DARPA, and in part by Terraswarm, an SRC
program sponsored by MARCO and DARPA.
Additionally, this material is based upon work supported by the National Science
Foundation Graduate Research Fellowship Program under grant numbers DGE-1256260
and DGE-1106400,
NSF/Intel CPS Security under grant 1505684,
and generous gifts from Intel.

We would also like to thank
our anonymous reviewers for their insightful feedback,
our shepherd Neal Patwari,
and the many collaborators that helped make this work possible.
Specifically Amit Levy and the Tock development team for their
support in using an under-development operating system,
Anthony Rowe, Craig Hesling, and Artur Balanuta for LoRaWAN gateway hardware and support,
William Huang for designing the environmental sensing module,
Yifan Hao for designing the radar module,
Theo Miller for porting the software library to Arduino,
Justin Hsieh for work on the communication protocol,
Ken Lutz for helping with deployment logistics,
and Noah Klugman for help with \name assembly.
          % 0.0 pg

\bibliographystyle{ACM-Reference-Format}
\bibliography{bib}

\end{document}